\documentclass[12pt,titlepage]{article}
\usepackage{JMR}

\usepackage{authblk}

\usepackage{color}
\usepackage{amsmath}    
\usepackage{subfigure}  

\newcommand{\bnabla}{\boldsymbol{\nabla}}
\newcommand{\bu}{\boldsymbol{u}}
\newcommand{\bk}{\boldsymbol{k}}
\newcommand{\bx}{\boldsymbol{x}}
\newcommand{\bp}{\boldsymbol{p}}
\newcommand{\bA}{\boldsymbol{A}}

\newcommand{\al}{\alpha}
\newcommand{\bet}{\beta}

\newcommand{\be}{\mathbf{e}}
\newcommand{\defn}{\stackrel{\text{def}}{=}}

\title{Refraction of swell by surface currents }
\author[1]{Basile Gallet}
\author[1]{William R. Young }
\affil[1]{ Scripps Institution of Oceanography\authorcr University of California at San Diego \authorcr La Jolla, CA 92093-0213 \authorcr
\tt{basile.gallet@gmail.com} and  \tt{wryoung@ucsd.edu}}

\date{}

\singlespace

\begin{document}
\maketitle

\begin{abstract}

Using recordings of swell from pitch-and-roll buoys, we have reproduced the classic observations of  long-range surface wave propagation originally made by Munk et al. (1963) using a triangular array of bottom pressure measurements. In the modern data,  the direction of the incoming swell fluctuates by about $\pm 10^\circ$ on a time scale of one hour. But if  the incoming direction is averaged over the duration of an  event then, in contrast with the observations by Munk et al. (1963), the sources inferred by great-circle backtracking  are most often in good agreement with the location of large storms on weather  maps of the Southern Ocean. However there are a few puzzling  failures of great-circle backtracking  e.g., in one case,  the direct  great-circle route is blocked by the Tuamoto Islands and  the inferred source falls on New Zealand. Mirages like this occur more frequently in the bottom-pressure observations of Munk et al. (1963), where several inferred sources fell on the Antarctic continent.


Using spherical ray tracing we investigate the hypothesis that the refraction of waves by surface currents produces the  mirages.  With reconstructions of surface currents inferred from satellite altimetry, we show that mesoscale vorticity significantly deflects  swell away from great-circle propagation so that the source and receiver are connected by a bundle of many rays, none of which precisely follow a great circle. The   $\pm 10^\circ$ directional fluctuations  at the receiver result from the arrival of wave packets that have travelled along the different rays within this multipath.   The occasional failure of great-circle backtracking, and the associated mirages,  probably results from partial topographic obstruction of the multipath, which biases the directional average at the receiver.

\end{abstract}

\section{Introduction}

Following the disruption of the  1942 Anglo-American landings on the Atlantic beaches of  North Africa  by six-foot surf \citep{Atkinson}, the forecasting of surface gravity waves became a wartime priority. These first  surface-wave forecasts  were based on weather maps, and the relatively predictable propagation of swell, and were used  to determine optimal  conditions for amphibious assault \citep{vonStorch,Bates}. Wartime work showed that  long surface waves,   generated in stormy regions of the globe, can  travel for many thousands of kilometers  before breaking on distant shores \citep{Barber}.   In the following  decades, Munk, Snodgrass and collaborators observed that  swell generated in the ocean surrounding   Antarctica travels half way around the Earth \citep{Munk57, Munk63,Snodgrass}. Thus, after a transit of 5 to 15 days, the waves created by winter storms in the Southern Ocean produce  summer surf in California.

\cite{Barber} used  linear wave theory  to relate the range of a distant storm to the rate of change of peak wave frequency at an observation point. But the  formula of Barber \& Ursell  --- see  (\ref{BU}) below --- provides no information about the direction of the source.  The first attempt at measuring the direction of the incoming swell was made by \citet{Munk63}  using  an array of three pressure-transducers on the sea bottom offshore of San Clemente Island. The method is analogous to astronomical interferometry and was used  to infer the direction of incoming wave-trains from June to October of 1959.  Assuming that swell travels on great-circle routes, these observers  combined Barber \& Ursell's estimate of the range with interferometric direction   to  locate  storms in the Southern Ocean. Although these inferred sources could be related to  storms on weather maps, the location errors for some events were as much as  $10^{\circ}$ of arc along the surface of the Earth, or $1000$km. Moreover, there is a systematic error in the observations of \citet{Munk63}: the inferred source is most often to the south of the actual storm \citep{Munkcomm}. Because of this southwards shift,  three of the thirty inferred storms even fell on the Antarctic continent, and several others on  sea-ice. The unavoidable conclusion is that  surface waves do not travel precisely on great-circle routes.  Effects such as planetary rotation \citep{Backus}, or  the slightly spheroidal figure of the Earth, were shown to be far too small to explain the observed departure from great-circle propagation. 
We now believe  that these mirages were caused by a combination of  topographic obstruction  (Munk 2013) and refraction by surface currents.

 \cite{Kenyon} advanced the hypothesis that the refraction of surface gravity waves by  major ocean currents, particularly the Antarctic Circumpolar Current,  might explain the discrepancies reported by \cite{Munk63}. A packet of surface waves  propagating through these currents is in a medium that varies slowly on the scale of a wavelength. The  currents induce a doppler-shift in the surface wave frequency, 
\begin{equation}
\Omega(\bx,\bk )=\sqrt{g k} + \bu(\bx)\cdot \bk \, , 
\label{dispersion}
\end{equation}
with $\bk$ the wavevector, $k = |\bk|$, $\bu$ the surface velocity, and $g$ the acceleration of gravity. Over many wavelengths refraction by the currents changes the direction of propagation of  a wave packet so that it departs from a great circle. Given the surface current $\bu(\bx,t)$, the ray  equations describe the coupled evolution of the local wavevector $\bk(t)$ and the position of the wave-packet $\bx(t)$, just as the propagation of light in a slowly varying medium follows  the rules of geometrical optics \citep{Whitham,Buhler,LLfluid}. Using an idealized model of the Antarctic Circumpolar Current as a parallel shear flow,  \cite{Kenyon}  solved the ray-tracing equations and found appreciable deflections for rays making a grazing incidence with the current. 

 
  

Here we re-visit Kenyon's hypothesis with the benefit of modern reconstructions of ocean surface currents and wave climate. 
In section \textbf{2} we analyze modern swell measurements from pitch-and-roll buoys (Kuik, van Vledder \& Holthuijsen, 1988). We observe that the direction of incident swell measured by a deep-water  buoy  close to San Clemente Island varies by $\pm 10^{\circ}$ on a time scale of hours. After averaging over these directional fluctuations  to determine the mean direction, we use great-circle backtracking and find that for most events there is excellent agreement between the location of the  inferred source and wave maxima on weather maps. But, in a few cases, great-circle backtracking places the wave source on land, as much as $10^{\circ}$ from the nearest storm on weather maps.  Section \textbf{3} develops the theory of spherical ray tracing, including the effects of surface currents.  In section \textbf{4} we compute ray paths through realistic surface currents and show that because of deflection by currents  the source and the receiver are connected by a bundle of rays (a ``multipath"). None of the rays in a multipath  follow the great circle.  In section \textbf{5}  we show that refraction by surface currents can explain both the magnitude of the  directional fluctuations at the receiver and their quantitative dependence on wave frequency.   We conclude in  section \textbf{6} by proposing a mechanism based on the interplay between surface currents and topography to explain the systematic shift towards the South and the frequent inference of sources on land in the observations by \citet{Munk63}.

 
\section{Modern data \label{mod}}

We have reproduced the observations of \cite{Barber} and  \cite{Munk63} using modern measurements provided by recordings of waves from pitch-and-roll buoys  deployed by NOAA's National Data Buoy Center (NDBC). Using three-dimensional accelerometers, these buoys measure the wave height, period and direction  \citep{Kuik}. The intensity of the swell can be characterized by the Significant Wave Height (SWH), defined as the average height of the highest one-third of the waves, and recorded by the station every hour during a 20-minute sampling period. The SWH is usually of the order of four times the root-mean-square surface elevation.

NOAA Station 46086 is located in the San Clemente basin at latitude $32.5^\circ$N and  longitude $118.0^\circ$W, close to where \citet{Munk63} used  bottom-pressure transducers to measure swell direction. The accelerometer buoy  at station 46086 has the great advantage of floating in $2000$ meters of water, so that  the influence of bathymetry is negligible. By contrast, the  station of \cite{Munk63} was in approximately 100 meters of water, and refraction by local bathymetry impacted the measured swell direction \citep{Munk2013}.

\begin{figure}
\begin{center}
 \subfigure[]{
\includegraphics[width=70 mm]{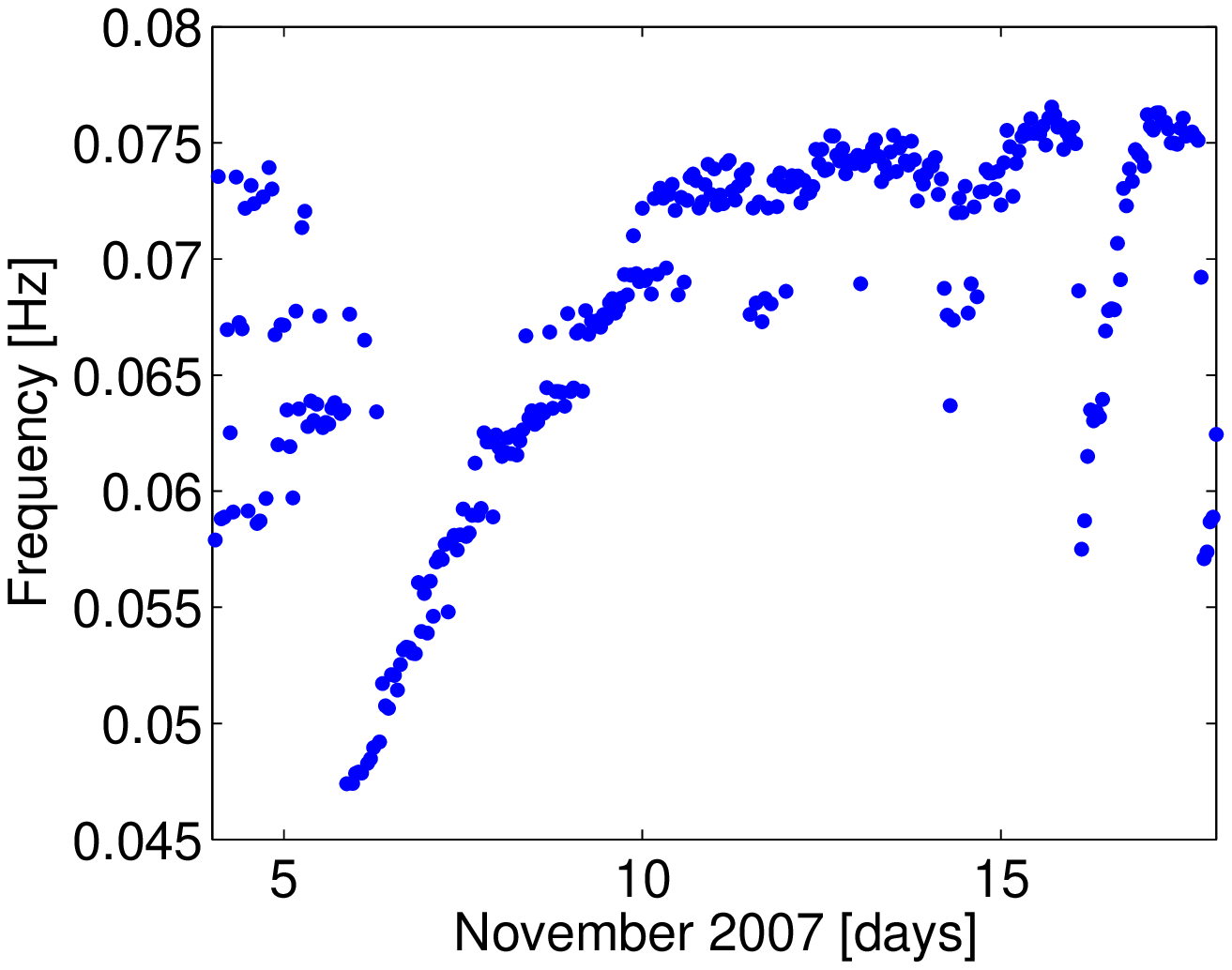}
}
 \subfigure[]{
\includegraphics[width=70 mm]{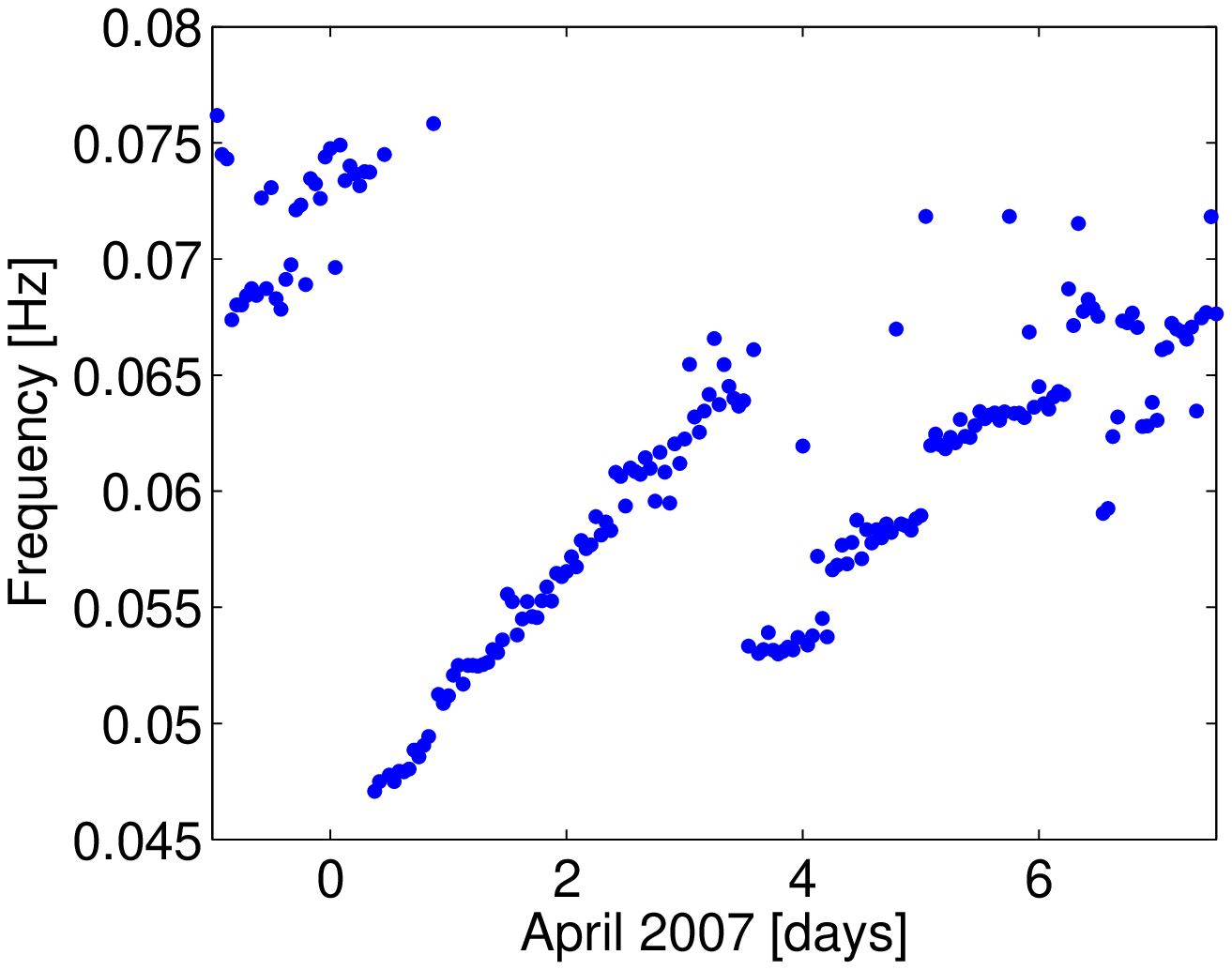}
}
\end{center}
\caption{Identification of two swell events recorded at NOAA Station 46086 in the San Clemente basin. The ``ramps" in a time-frequency plot signal the arrival of swell from a distinct,  distant source. Panel (a) shows the peak spectral frequency as a function of time  recorded for 14 days in  November 2007; we study the event occurring between the 6th and 10th of November. Panel (b) shows the peak frequency as a function of time  recorded for 7 days in  April 2007; we study the event between the 31st of March and the 3rd of April.}
\label{LongerSwell} 
\end{figure}

The spectral intensity is provided in frequency bins spaced by $5$mHz in the low-frequency range, together with the mean direction of the signal for each one of these bins. These directional spectra are based on an observation length of  twenty minutes and are delivered every hour. The arrival of  swell from a distinct source at  buoy 46086 is indicated by a strong spectral peak  that shifts towards higher frequencies as time goes on, with a timescale of several days: see Fig. \ref{LongerSwell}. The progressive shift of the spectral peak to higher frequencies  is caused by the dispersive propagation of surface waves: longer waves travel faster, so that  fast low-frequency  swell reaches the receiver first \citep{Barber}.

To isolate swell events, we track the peak frequency of the low-frequency part of the spectrum. To smooth  the discretization from frequency bins, we first locate the low-frequency maximum in spectral intensity, before performing a weighted average of this frequency with the frequencies of the two neighbouring bins, using weights equal to the spectral intensity in each of the three bins.

From each directional spectrum, we extract the incident angle of the swell, which we define as the mean direction for the frequency bin with maximum spectral intensity: this incident angle is computed every hour using 20-minute long samples.

\subsection{A case study: the event of November 2007}

\begin{figure}
\begin{center}
 \subfigure[]{
\includegraphics[width=70 mm]{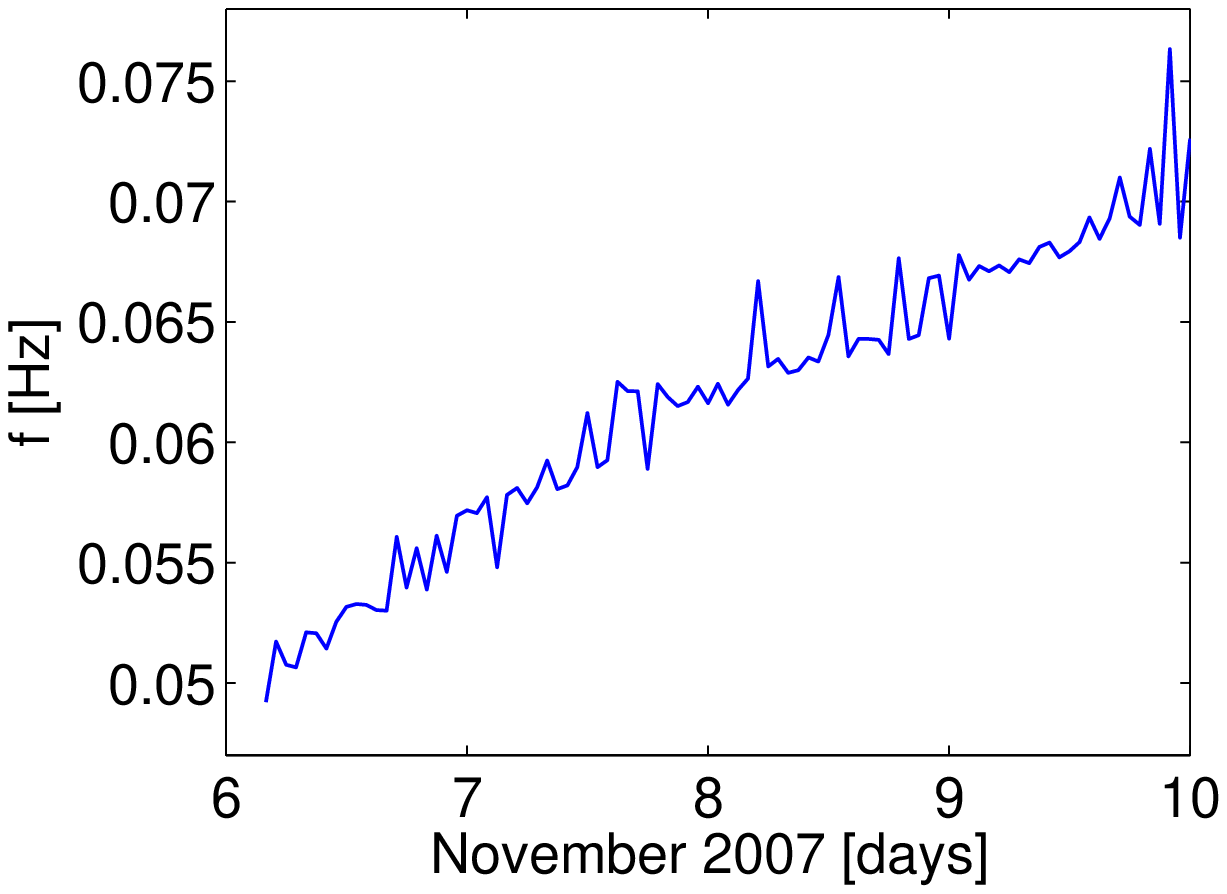}
}
 \subfigure[]{
\includegraphics[width=70 mm]{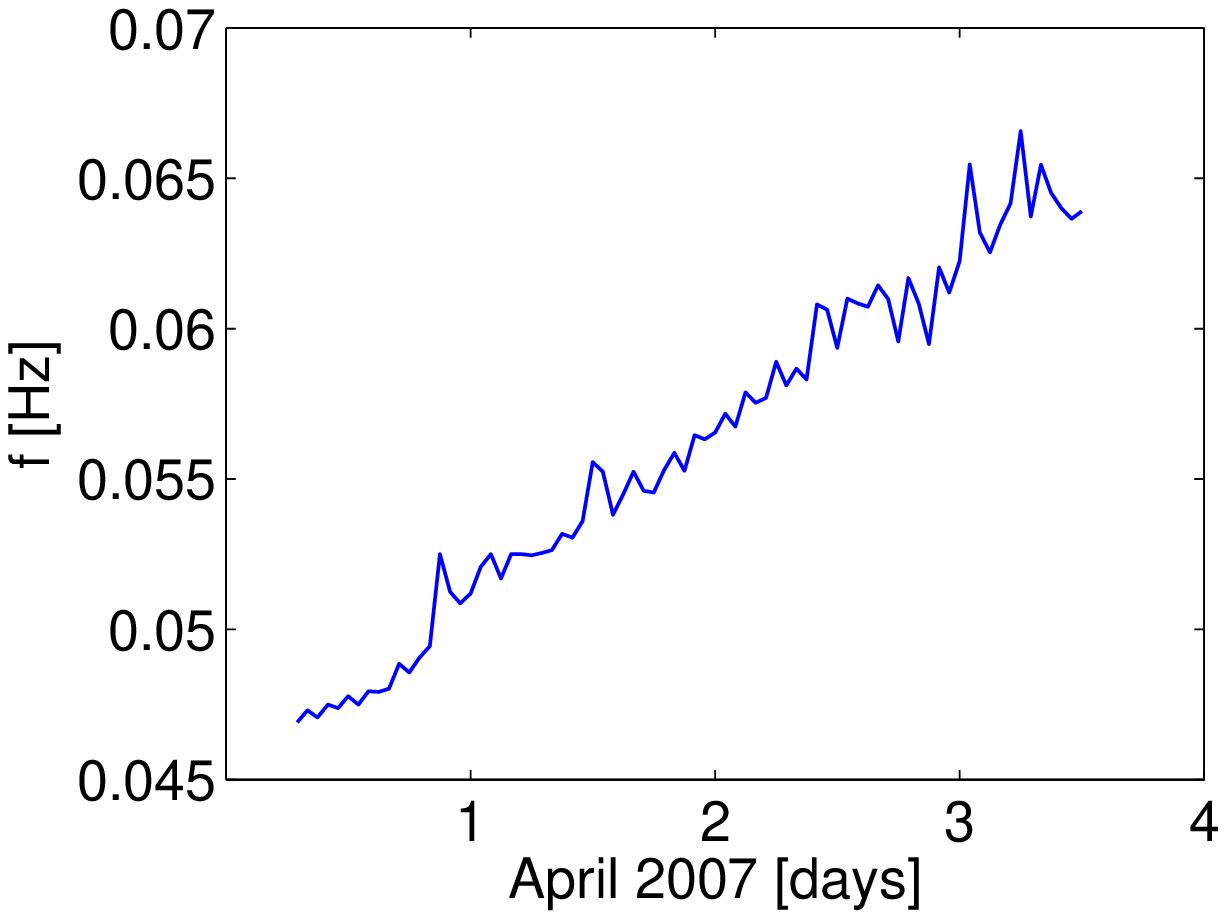}
}
 \subfigure[]{
\includegraphics[width=70 mm]{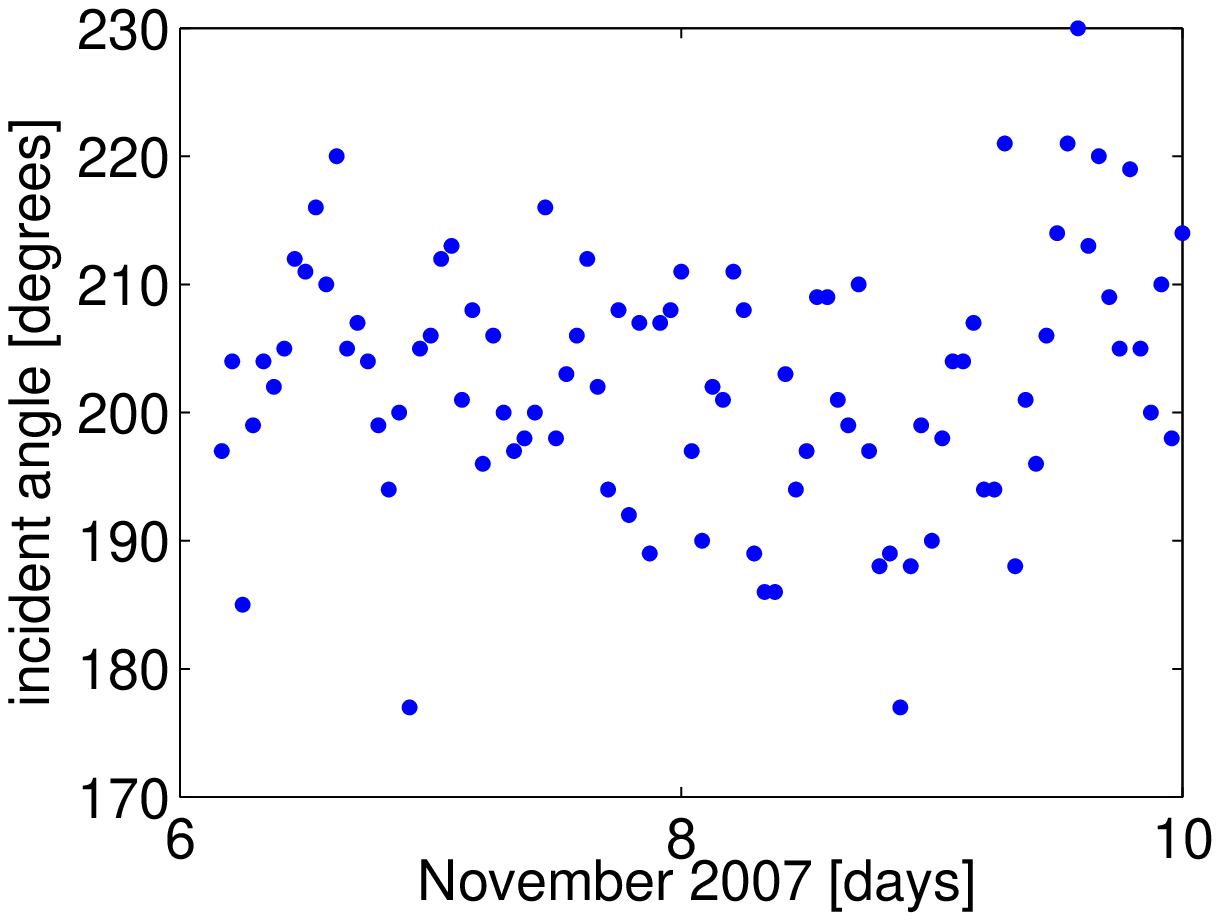}
}
 \subfigure[]{
\includegraphics[width=70 mm]{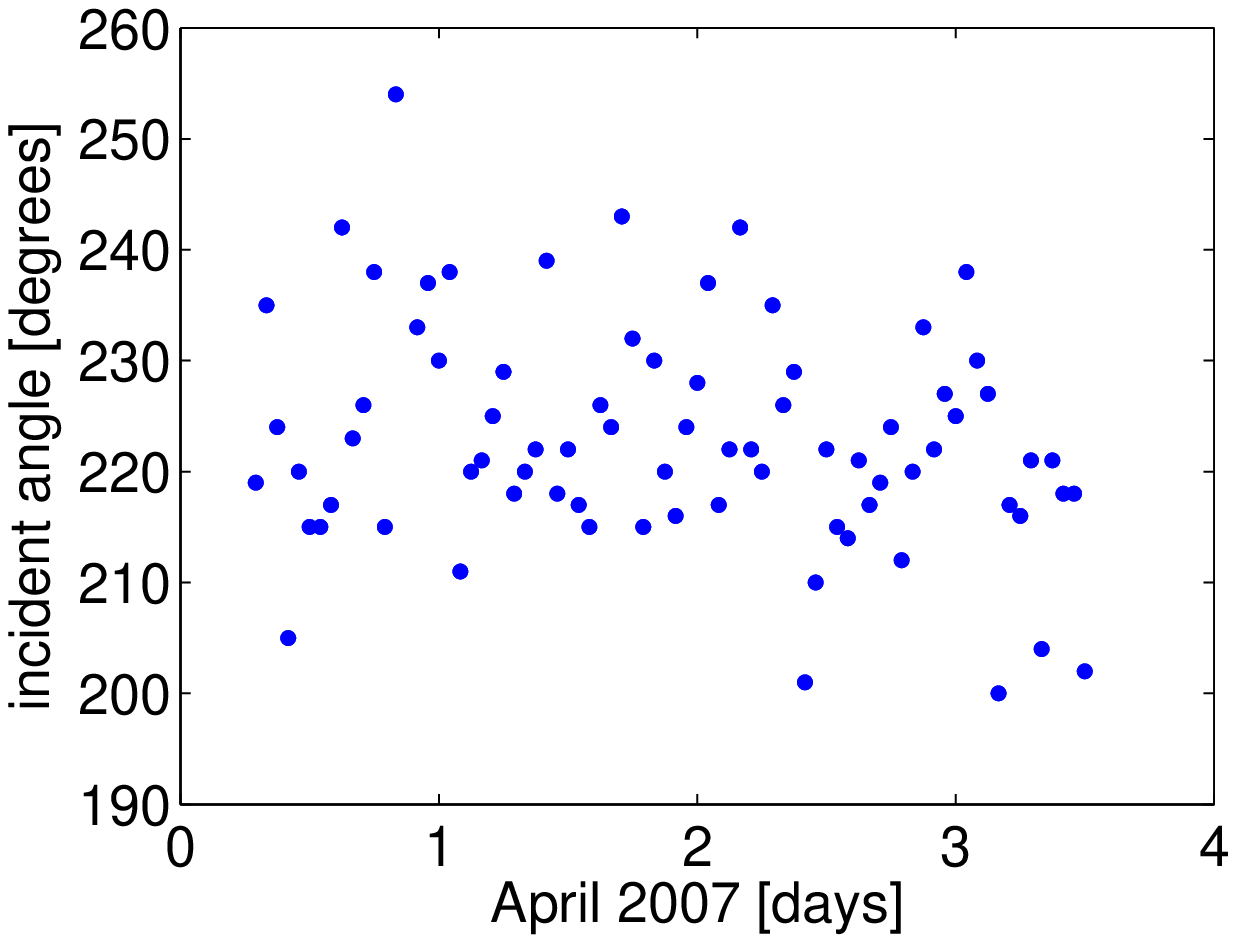}
}
 \subfigure[]{
\includegraphics[width=70 mm]{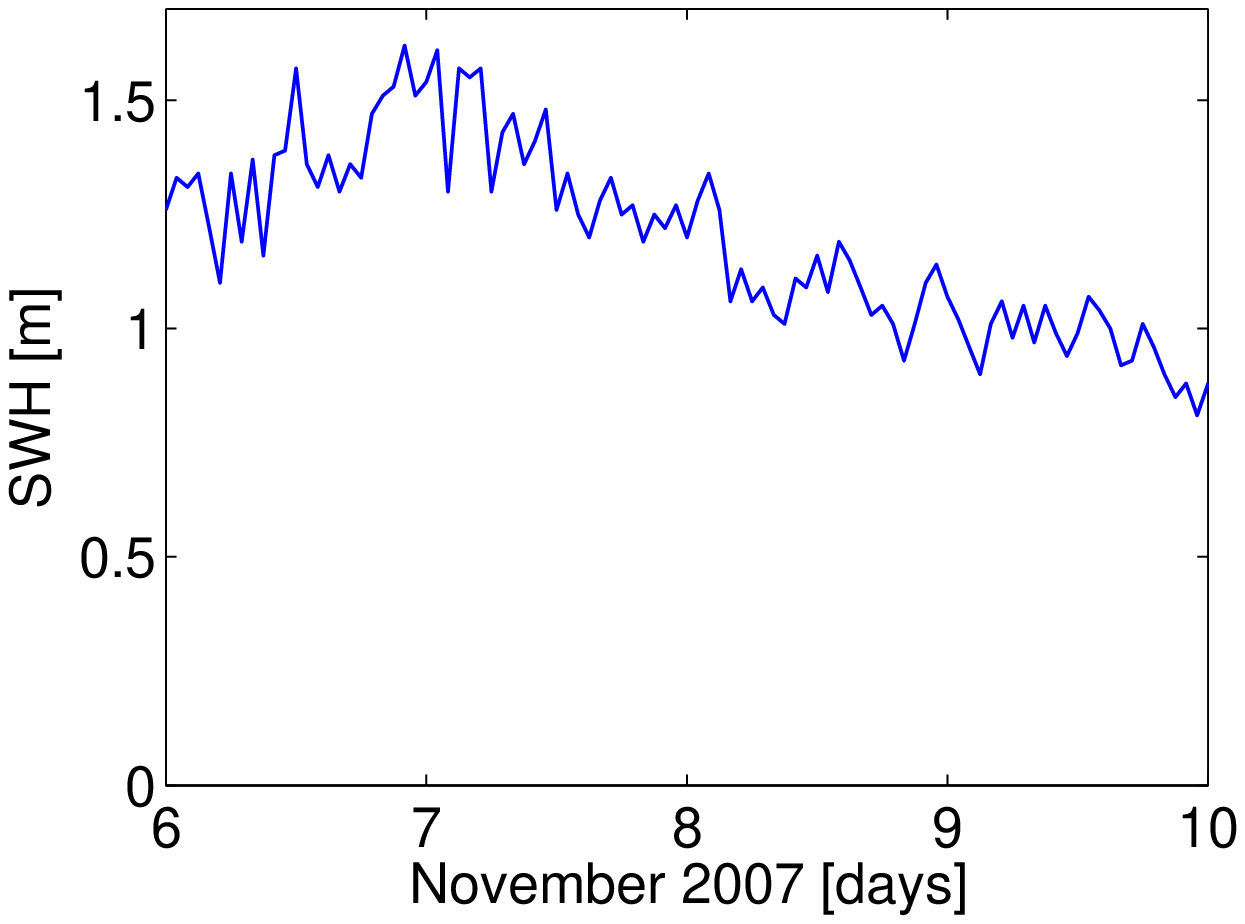}
}
 \subfigure[]{
\includegraphics[width=70 mm]{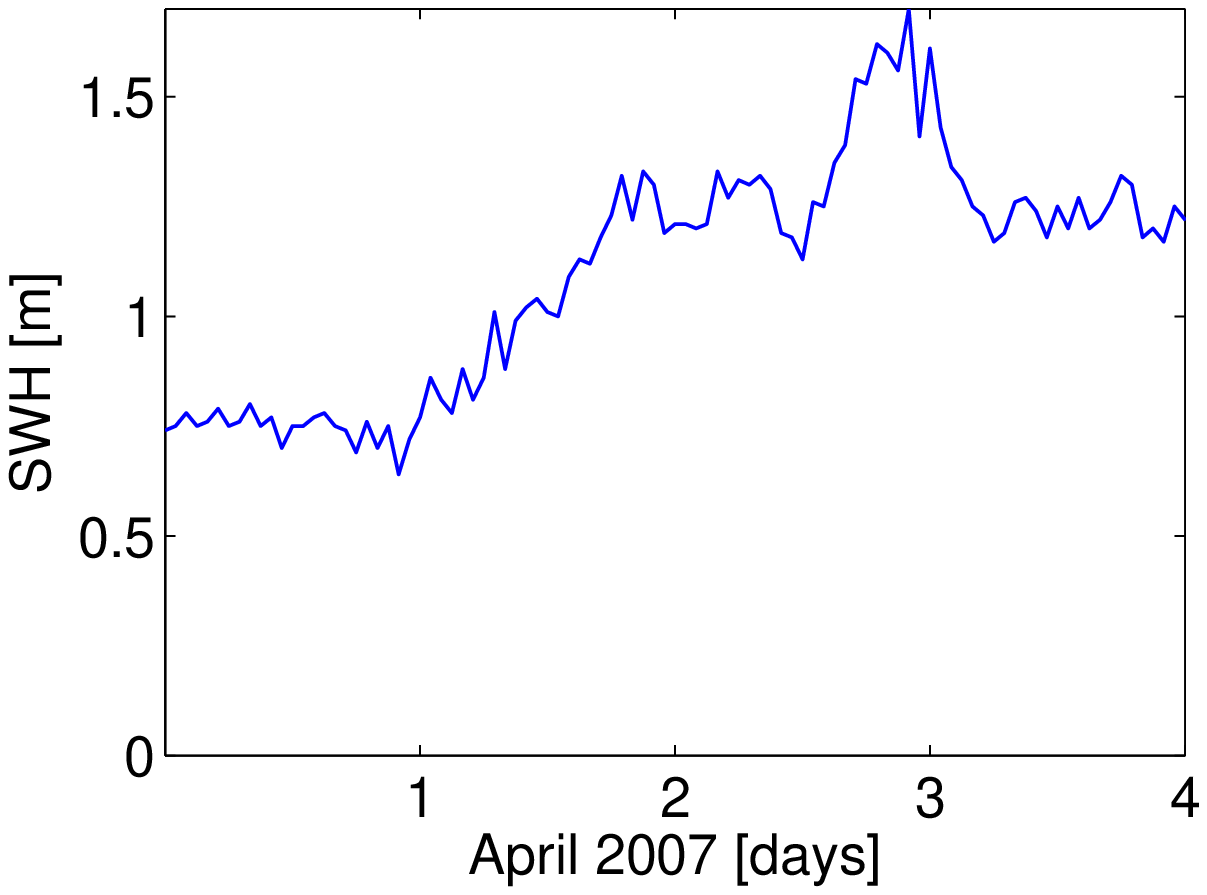}
}
\end{center}
\caption{Two swell events recorded at NOAA Station 46086 in the San Clemente basin. Panels (a), (c) and (e) correspond to a swell event recorded in November 2007, and panels (b), (d) and (f) to a swell event recorded in April 2007. The top panels show the estimated peak frequency $f$ as a function of time. $f$  increases linearly with time due to dispersive propagation of surface waves. The middle panels show the incident angle measured clockwise from North by the buoy. The incident angle fluctuates around a mean value of $204^\circ$ for panel (c) and $224^\circ$ for panel (d). The bottom panels show the SWH in meters.}
\label{2swellevents} 
\end{figure}

The left-hand side panels of Fig. \ref{2swellevents} focus  on the particularly striking  event in November 2007: there are four consecutive days of swell with SWH at around $1.2$ meters in  panel (e). Fig. \ref{2swellevents}(a) shows the wave propagation diagram. As first observed by Barber \& Ursell, the peak frequency increases linearly with time and, using $\omega=\sqrt{gk}$, one can infer the distance $L$ between the source (the storm) and the receiver (buoy 46086) from the slope of the frequency versus time on the wave propagation diagram:
\begin{equation}
\frac{df}{dt}=\frac{g}{4 \pi L}\, .
\label{BU}
\end{equation}
Above, $f=\omega/(2 \pi)$ is the frequency in Hertz. Using a linear fit to the early part of the swell event reported in figure \ref{2swellevents}, we find a distance of $86.7^\circ$ of arc between the storm and buoy 46086. The intersection of the fitting line with the $f=0$ axis gives the date of birth of the storm which is October $29$th 2007, at around 23:00 GMT.


 To infer the direction of the source we turn to the incident-angle signal in  Fig. \ref{2swellevents}(c). There are $\pm 10^{\circ}$ fluctuations in the direction of the incident waves at the buoy.  We remove these fluctuations by averaging, and so find that the wave signal at station 46086 comes from $204^{\circ}$, measured clockwise from North. We hypothesize  that the $\pm 10^{\circ}$ directional fluctuations in Fig. \ref{2swellevents}(c) are too large to be simply instrumental noise, and that there might be physical information in the directional measurements. We return to further  discussion of this point in section \textbf{5} e.g., see Fig. \ref{thetarmsvsf_allrays} and the supporting discussion. 

\begin{figure}
\begin{center}
\includegraphics[width=80 mm]{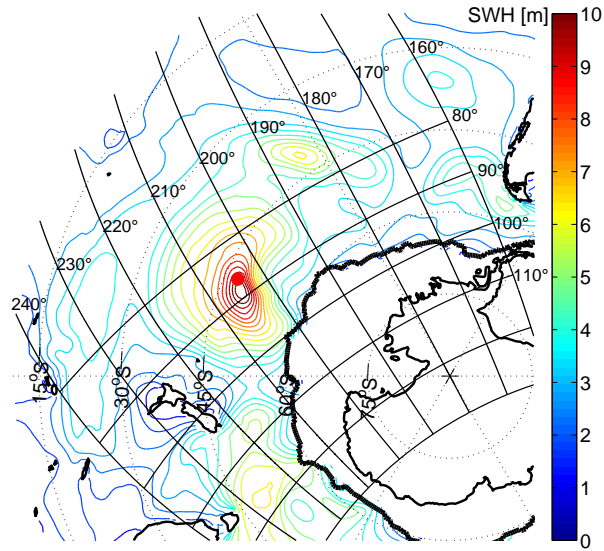}\\
\end{center}
\caption{Color contours indicate significant  wave height (SWH) in meters from the ECMWF ERA reanalysis on October 30th, 2007, at midnight GMT, shown using a South Polar projection. The thick black line is the sea-ice limit. The solid black grid shows great-circle routes from the NOAA station 46086, and lines of constant range from this station.  The red spot,  which is very close to the region of maximum SWH, indicates the source inferred from swell recorded at Station 46086.}
\label{October30} 
\end{figure}

The inferred range and direction locate the wave source on  weather maps that are made available by the European Center for Medium-Range Weather Forecasts (ECMWF). The ECMWF interim reanalysis \citep{ERAInterim} provides sea-ice cover, 10-meter wind speed, and also SWH computed using the WAM wave model and assimilation of altimeter data \citep{WAM,KomenBook}. We identify the southern sources of swell, corresponding to  southern-ocean storms,  as  strong local maxima in 10-meter wind speed, and  as large  SWH.  In Fig. \ref{October30} we compare the relevant ECMWF SWH with the source inferred from the accelerometer buoy recordings (the red dot).  The buoy data analyzed above predict a storm at a range of $86.7^\circ$ on a great-circle route making an angle of $204^\circ$ going clockwise from North at the buoy, on October $29$th 2007 at 11 pm. The inferred location, shown as a red dot in  Fig. \ref{October30}, is in excellent  correspondence with the  SWH maximum. In the example  of Fig. \ref{October30},  great-circle backtracking  works very well: refraction by surface currents and topographic effects do not spoil the inference of the source. 
 
%
%
%
\subsection{Another case study: the event of April 2007}

\begin{figure}
\begin{center}
\includegraphics[width=80 mm]{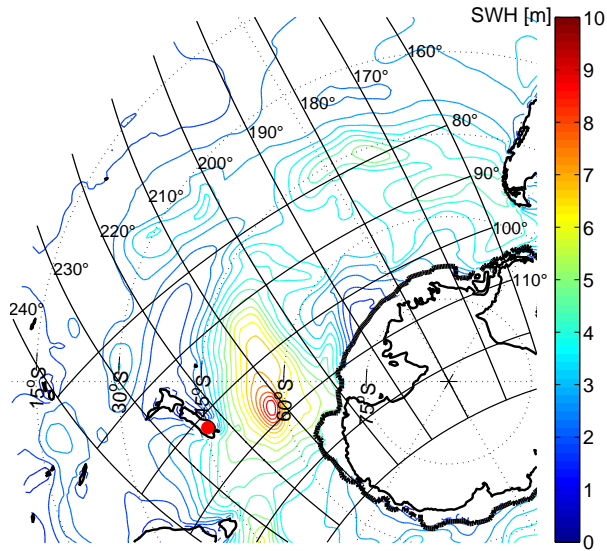}\\
\end{center}
\caption{A mirage: the swell recorded at NOAA station 46086 seems to originate from New Zealand.  The inferred location (the red dot) is displaced  by approximately $10^\circ$ of arc from the region of maximum SWH. Color contours indicate SWH  in meters for the ECMWF ERA reanalysis for  the storm of March 23rd, 2007 at 6am GMT. The solid black grid shows great-circle routes from the NOAA station 46086, and lines of constant range from this buoy.}
\label{March23} 
\end{figure}

The agreement shown in Fig. \ref{October30} is the most frequent situation we found in our analysis of the data from accelerometer buoys: for 14 of the 18 swell signals we analyzed, the inferred source corresponds to a maximum in ECMWF surface wave height within $5^\circ$ of arc. But we also found a few examples for which the swell signal was very clean e.g., Fig. \ref{LongerSwell}(b), yet great-circle backtracking  resulted in a bad estimate of source location. In the right-hand panels of Fig. \ref{2swellevents}, and in Fig. \ref{March23} we focus on this event in April 2007. The SWH at station 46086 is of the order of 1 meter and the frequency of the swell increases linearly with time. Yet in Fig. \ref{March23} the direction of the inferred source is  approximately $10^\circ$ from the maximum in ECMWF surface wave height. Moreover, the $10^\circ$ error puts the  inferred source of this April 2007 event  on New Zealand. This mirage is reminiscent of the Antarctic sources inferred by \cite{Munk63}. 

Mirages seem to occur preferentially when there is shallow topography, or even land, close to the great-circle route between the storm and the receiver. In the case of Fig. \ref{March23}, a dense part of the Tuamotu Archipelago between  $14^{\circ}$ to  $18^{\circ}$ South and  $148^{\circ}$ to $140^{\circ}$ West ---see Fig. 50 of \cite{Munk63}  --- blocks the wave packets propagating on the great-circle route between the storm and the receiver. Only rays that are deflected strongly enough by surface currents to go around this Tuamotu blockage can reach the receiver. In anticipation of results from section \textbf{4}, a  possibility is that because of the distribution of surface currents,  some wave-packets were deflected west of the Tuamotu blockage, so that the inferred source appears on New Zealand: the mirage in Fig. \ref{March23}  results from the interplay between Tuamotu blockage and refraction by surface currents.

\section{Waves on the surface of a sphere}

Modeling the results of section \textbf{2} requires tracing rays  on the  surface of a sphere, including the effects of refraction by surface currents. \cite{Backus} developed this ray theory, including shallow-water effects and rotation, but without considering ocean currents. In this section we provide an account of the relevant  theory required for the model in section \textbf{4}.

\subsection{The ray equations in spherical coordinates}

Let us denote the phase of a wavepacket by $S(\bx,t)$. Then the  frequency and local wavevector are respectively $-\partial_t S$ and $\bk={\boldsymbol {\nabla}} S$.  The dispersion relation can be written in terms of $S(\bx,t)$ as  $\partial_t S+ \Omega(\bx,\bnabla  S)=0$. This partial differential equation is the Hamilton-Jacobi equation for a mechanical system with action $S$ and Hamiltonian $\Omega$. The solution  is accomplished via Hamilton's equations, which in this context are also called the ray equations.  These are evolution equations for the position $\bx(t)$ and wavevector $\bk(t)$ of the wave-packet \citep{Buhler}. 

For the spherical problem at stake, we use latitude $\psi$ and longitude $\phi$, with unit vectors $\textbf{e}_\psi$ and $\textbf{e}_\phi$. The conjugate momenta are then $p_\psi=\partial_\psi S$ and $p_\phi=\partial_\phi S$. Using the expression for the gradient in terms of latitude and longitude, the wavevector is 
\begin{equation}
\bk= \frac{p_\psi}{R} \be_\psi + \frac{p_\phi}{R \cos \psi} \be_\phi \, ,
\end{equation}
with $R$ the radius of the Earth. The current velocity is $\bu= u(\psi,\phi)  \be_\phi + v(\psi,\phi)  \be_\psi$,  and   (\ref{dispersion}) then gives the Hamiltonian in terms of $\psi$, $\phi$ and their conjugate momenta:
\begin{equation}
\Omega(\psi,\phi,p_\psi,p_\phi) = \sqrt{\frac{g}{R}}\,  p^{{1}/{2 }}   +  p_\psi \, \frac{v(\psi,\phi)}{R}+  p_\phi\,  \frac{u(\psi,\phi)}{R \cos \psi}   \, ,
\end{equation}
where
\begin{equation}
p\defn \left( p_\psi^2 +\frac{p_\phi^2}{\cos^2 \psi }\right)^{1/2}\, .
\end{equation}
The spherical ray equations are then obtained from $\Omega(\psi,\phi,p_\psi,p_\phi)$  via:
\begin{eqnarray}
 \dot{\psi}  =  \partial_{p_\psi} \Omega & = & \sqrt{\frac{g}{R} }\,  \frac{p_\psi}{2 p^{{3}/{2}}  } +\frac{v}{R} \, , \label{ray1}   \\
 \dot{\phi}  =  \partial_{p_\phi} \Omega & = &  \sqrt{\frac{g}{R} }\,  \frac{ p_\phi}{2 p^{3/2} } \,  \frac{1}{\cos^2 \psi} +\frac{u}{R \cos \psi}\, ,  \label{ray2}   
  \end{eqnarray}
  and
  \begin{eqnarray}
 \dot{p_\psi}  =  - \partial_{\psi} \Omega & = & - \sqrt{\frac{g}{R} } \, \frac{ p_\phi^2 }{2 p^{3/2} }\, \frac{\sin \psi}{\cos^3 \psi} - \frac{p_\psi }{R} \, \partial_\psi v - \frac{p_{\phi}}{R}\,  \partial_\psi \frac{u}{\cos \psi} \, , \label{ray3}\\
 \dot{p_\phi}  =  - \partial_{\phi} \Omega & = & -\frac{p_\psi }{R}\, \partial_\phi v -\frac{p_\phi}{R} \frac{ \partial_\phi u}{ \cos \psi}\, .\label{ray4}
\end{eqnarray}
An alternate method for deriving ray equations in spherical geometry is given in \cite{Hasha}. In the special case of propagation through a still ocean, $\bu=0$,  the conjugate momentum $p_{\phi}$ is constant and equations \eqref{ray1} through \eqref{ray4}  reduce to those of \cite{Backus} and describe great circle propagation.

\subsection{A special solution}

\begin{figure}
\begin{center}
\includegraphics[width=50 mm]{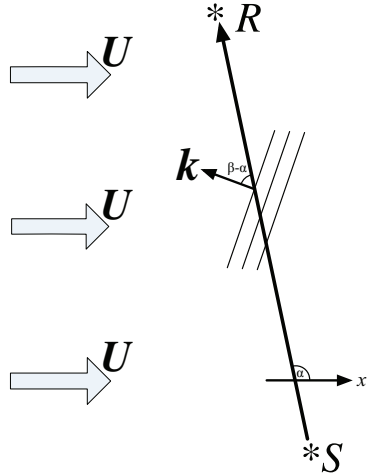}\\
\end{center}
\caption{
A ray from a source $S$ to a  receiver $R$ is not bent by a uniform current $U$. The  wave vector $\bk$ is inclined to the ray-path so that part of the group velocity compensates for advection by the current. For clarity, this schematic shows a large value of the angle $\beta-\alpha$ between  the ray $RS$ and  $\bk$. Realistic surface currents are weak compared to the group velocity $v_g$ and hence $\beta-\alpha$ is at most one degree.}
\label{StraightRay} 
\end{figure}

An educational solution of the ray equations is obtained by considering the Cartesian case with a  uniform current $U$ flowing along the axis of $x$: see Fig. \ref{StraightRay}.  If the source $S$ is at the origin, and the receiver  $R$ is at  $\bx_R = r \cos \al \, \be_x+ r\sin \al \, \be_y$, then   the ray connecting $S$ to $R$ is a straight line, despite the Doppler shift corresponding to $U$. Thus, in planar geometry, a uniform current does not bend rays.

This straight-line propagation, while simple in principle, is perhaps  counterintuitive. Thus it is worthwhile to understand straight-line propagation through a uniform current  by  explicit solution of the ray equations. In cartesian geometry, the dispersion relation is $\Omega = \sqrt{g k} + U k_x$, with wave vector $\bk = k_x \be_{x}+ k_y \be_y $ and total wavenumber $k =\sqrt{k_x^2+k_y^2}$. The cartesian ray equations are 
\begin{equation}
\dot x = \partial_{k_x}\Omega =U+  \frac{1}{2} \frac{k_x}{k}  \sqrt{\frac{g}{k}}\, \qquad \qquad  \dot y = \partial_{k_y} \Omega = \frac{1}{2} \frac{k_y}{k}  \sqrt{\frac{g}{k}}\, ,
\label{ray666}
\end{equation}
and
\begin{equation}
\dot k_x = - \partial_x \Omega =0\, , \qquad \qquad \dot k_y = - \partial_y \Omega =0\, .
\end{equation}
The wave numbers $k_x$ and $k_y$  are  constant, and the solution of \eqref{ray666} is therefore
\begin{equation}
x = \left( U+  \cos \bet \, v_g \right)\, t  \, , \qquad \qquad y = \sin \bet\,  v_g\,  t\, ,
\label{ray667}
\end{equation}
where $(k_x \, , k_y) =   k (\cos \bet \, ,  \sin \bet)$ and $v_g = \sqrt{g/4k}$ is the group velocity.
Eliminating $t$ between $x$ and $y$ in \eqref{ray667}, and requiring that the ray pass through the receiver at $\bx_R$, one finds that the direction, $\bet$, of the wave vector $\bk$ is given by
\begin{equation}
v_g \sin (\bet - \al ) =  U \sin \al\, .
\label{ray668}
\end{equation}
The relevant case in oceanography is $U \ll v_g$, so that \eqref{ray668} can always be solved for $\bet$. In Figure \ref{StraightRay} the slight inclination of the wave vector $\bk$ to the straight-line ray path $SR$ is the small angle $\bet - \al$.

  The importance of this simple solution is that it shows there is no relation between ray bending and the \textit{speed} of currents.

The difference between $\al$ and $\bet$ is a previously unremarked source of error for directional inferences which  suppose that the wave vector $\bk$ is precisely parallel to the direction of propagation e.g., as assumed by us in section \textbf{2}, and previously by \cite{Snodgrass} and \cite{Munk63}.  With a typical velocity of oceanic surface currents $U=0.3$ m/s and swell with $500$ m wavelength, we obtain $U/v_g \simeq 0.02$, hence $\al-\bet$ is of the order of $1^\circ$: the direction $\beta$ of the wave vector $\bk$ is therefore a good, but not perfect, estimate of the direction of the straight ray between $S$ and $R$. By contrast, in the following we show that non-uniform currents induce ray-bending, resulting in the direction of the wave vector $\bk$ at the receiver being a poor estimate of the direction of the source, with errors often larger than $10^\circ$.

\subsection{Ray bending and the vertical vorticity of currents}

Returning to the spherical case, in the absence of currents the solutions of \eqref{ray1} through \eqref{ray4} are great-circle geodesics (analogous to the straight line in Fig. \ref{StraightRay}) connecting the source to the receiver. Non-uniform currents will refract or bend the rays away from great-circle paths. In fact,  it is the vertical vorticity of currents that is crucial for bending rays away from great circles \citep{Kenyon,Dysthe,LLfluid}.

The connection between ray bending and vertical vorticity is simplest in the case when waves travel much faster than currents. For example,  swell with a wavelength $500$m has a group velocity $v_g= \sqrt{g/4k}$ of $14$m s$^{-1}$, which is much faster than the  velocity of typical surface currents (at most $1$m s$^{-1}$). In this limit of fast wave-packets, the ray equations \eqref{ray1} through \eqref{ray4} reduce to the  simpler and more insightful curvature equation 
\begin{equation}
\chi \simeq \frac{\xi}{v_g} \, \label{Dysthe} \, ,
\end{equation}
which is valid to first order in $|\bu|/v_g$. Above, $\xi$ is the vertical vorticity of the current,\begin{equation}
\xi(\bx)  \defn \frac{  \partial_{\phi} v - \partial_{\psi} ( u \cos \psi)   }{R\cos \psi } \, ,
\end{equation}
and $\chi$ is the geodesic curvature: $\chi$  is the curvature of the trajectory projected onto the local horizontal plane. A curve with zero geodesic curvature is a great circle in the present context.

Rays are thus deflected by  vorticity just as  the horizontal trajectory of a charged particle is bent by a vertical magnetic field: vorticity is analogous to the magnetic field. 
The magnetic-field  analogy is the starting point of an alternate derivation of equation (\ref{Dysthe}): in the limit of slow currents, the magnitude of $\bk$ varies very little, and the group velocity of the waves remains almost constant, at the initial value $v_{g (0)}$. Let us approximate the square root in the dispersion relation (\ref{dispersion}) by the parabola which has the same slope at the initial value of $k$:
\begin{equation}
\Omega(\bx,\bk)  \simeq   2 \frac{v_{g (0)}^3}{g} |\bk|^2 + \bu \!\cdot\! \bk + \text{constant .} 
\label{approxdisp}
\end{equation}
This quadratic expression is the same as the quadratic Hamiltonian for a negatively charged particle in a weak magnetic field ${\bf \nabla} \times \bA(\bx)$, where $\bA(\bx)$ is the vector potential. In dimensionless form, the Hamiltonian is:
\begin{align}
\mathcal{H}(\bx,\bp)  &= \tfrac{1}{2} |\bp+\bA|^2 + \tfrac{1}{2} |{\bf \nabla} \times \bA|^2 \, , \\
&\simeq \tfrac{1}{2} |\bp|^2  + \bA\!\cdot\! \bp+\mathcal{O}(\bA^2) \, , 
\end{align}
where $\bp$ is momentum. The last expression corresponds to the weak-field limit.
The charged particle experiences a Lorentz force, and its trajectory is well-known to have a local curvature proportional to the strength of the magnetic field $\chi = |\bnabla \times \bA|/| \bp |$ \citep{Jackson}. For the wave-problem governed by the approximate dispersion relation (\ref{approxdisp}), this translates into ray curvature $|\bnabla \times \bu|/v_{g (0)}$, which is the relation (\ref{Dysthe}).

The significance of \eqref{Dysthe} is that  looking at a map of surface vorticity, one can assess which features will strongly deflect  swell, and in which direction the rays will bend. The result also shows that wave propagation through this moving medium is isotropic, despite the direction determined by the velocity field $\bu = u(\psi,\phi)  \mathbf{e}_\phi + v(\psi,\phi)  \mathbf{e}_\psi$. We use (\ref{Dysthe}) to understand and interpret the results obtained by integration of the exact ray equations (\ref{ray1}) through (\ref{ray4}).

\section{Deflection of swell by surface currents}

Maps of surface currents $\bu(\bx,t)$ are made available by satellite altimetry and scatterometry: the Ocean Surface Current Analysis in Real-time (OSCAR) dataset gives the surface-current velocity field with a spatial resolution of one-third of a degree and a temporal resolution of 5 days \citep{OSCAR}. OSCAR estimates the velocity $\bu(\bx,t)$   required to determine the trajectory of swell by  integration of the ray equations \eqref{ray1} through \eqref{ray4}. We interpolate the OSCAR surface velocity field linearly onto a triangular mesh, before integrating the ray equations (\ref{ray1}) through (\ref{ray4}) with a forward Euler method.

One-third of a degree does not fully resolve  mesoscale vorticity and thus computations based on OSCAR  underestimate the deflection of  gravity waves by currents. This underestimate indicates unambiguously that surface-current refraction is quantitatively sufficient to explain fluctuations of  $\pm10^\circ$ in the direction of the incoming swell shown in Fig. \ref{2swellevents}(c) and \ref{2swellevents}(d).

\subsection{A point source in the Southern Ocean}


\begin{figure}
\begin{center}
\includegraphics[width=120 mm]{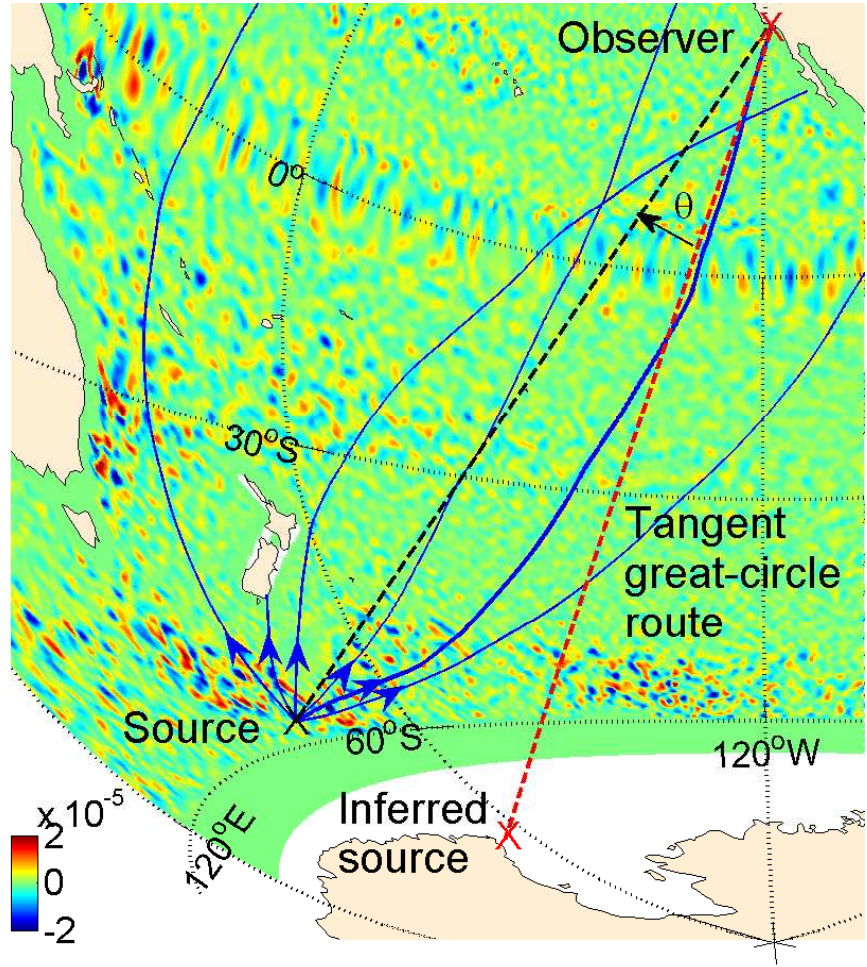}\\
\end{center}
\caption{Swell with $\lambda= 500$m emitted by a point-source in the Southern Ocean.  This figure uses an azimuthal equidistant projection so that great circles passing through the receiver (and only these) appear as straight lines i.e., the dashed straight lines are great circles. The color scale shows vertical  vorticity of OSCAR surface currents, $\xi$ in s$^{-1}$. The thick-curve is a ray that connects the source to the receiver. The thin curves are five other rays. Using great-circle backtracking (red dashed line),  the inferred source is far from the true source. }
\label{fig5}
\end{figure}

To model the observations of section \textbf{2},  we begin by  considering  a point-source in the Southern Ocean, at latitude $58^\circ$S and  longitude $160^\circ$E. This  source  emits surface waves in every direction, and with different wavelengths. For a given initial wavelength, we  numerically  integrate the exact ray-tracing equations \eqref{ray1} through \eqref{ray4}, starting from the source point and with different initial orientations of the wavevector $\bk$. We assume that the emission is  isotropic and so we shoot rays with the initial direction of $\bk$ uniformly distributed on the unit circle, with a $3\times 10^{-4}$ radian step (approximately one minute of arc). Six rays   determined by this procedure are shown in Fig. \ref{fig5}.  Out of all the emitted rays, an observer  receives only the few rays which connect the source to the receiver. Thus we  keep rays which reach the receiver within a radius of 30 nautical miles. The thick curve in Fig. \ref{fig5} is an example of such a ray.  Because the ray connecting the source to the receiver in Fig. \ref{fig5} is not a straight line we conclude that  swell is significantly deflected from a  great-circle route by the OSCAR currents. For  rays incident on the  receiver, we define the deflection angle $\theta$ as the angle between the great-circle route and the direction of the ray at arrival. We use the convention $\theta<0$ if the ray at arrival is South of the great-circle route between source and receiver.


If an observer  backtracks along the great circle indicated by the direction of arrival of the thick ray in Fig. \ref{fig5}, and  determines  the range  with \eqref{BU},  then the inferred source is over sea-ice and almost on Antarctica. In other words, refraction by OSCAR currents has produced a mirage.

\begin{figure}
\begin{center}
\includegraphics[width=100 mm]{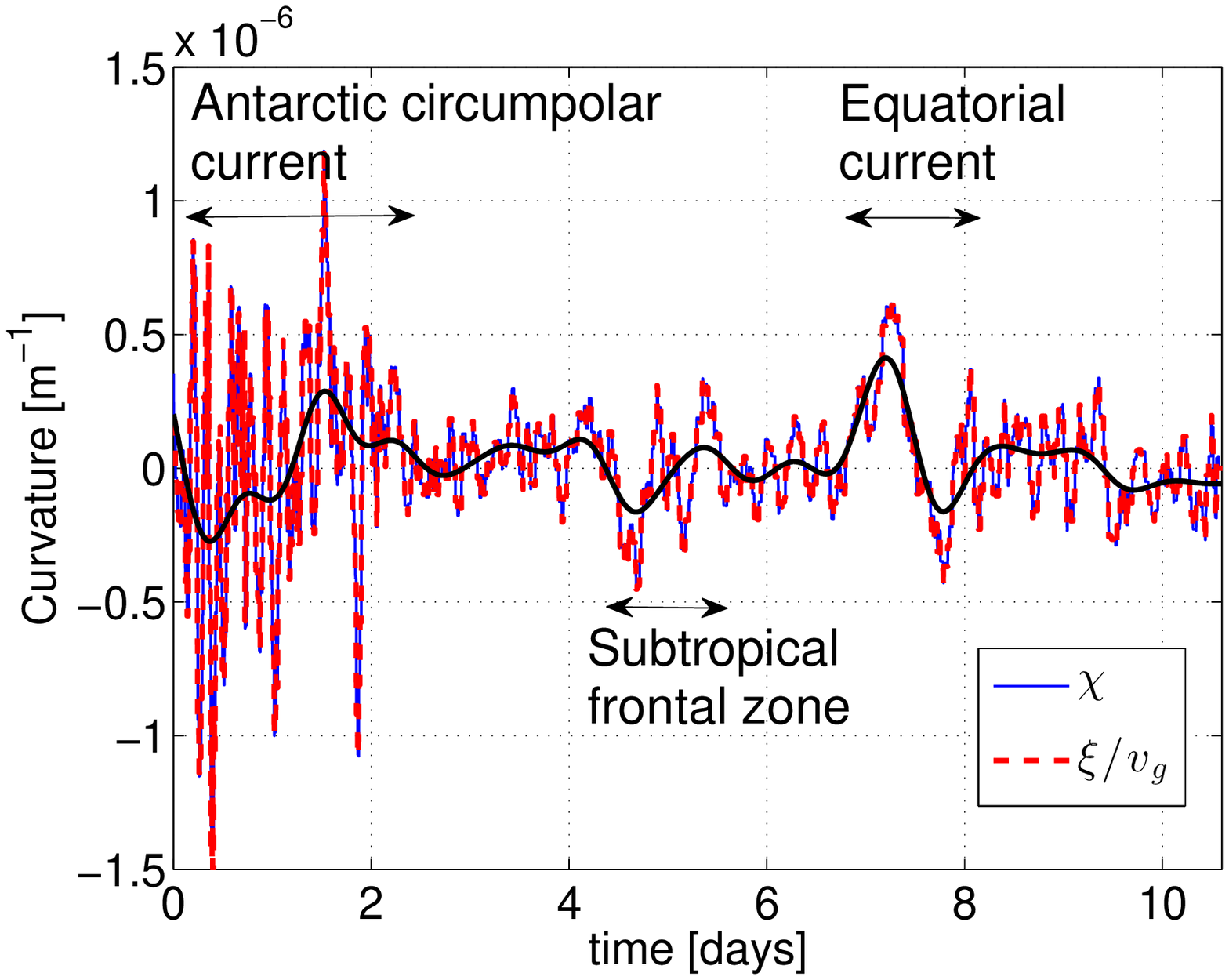}
\end{center}
\caption{ For the thick ray in Fig. \ref{fig5} the ray curvature (blue)  is almost equal to the surface vorticity divided by group velocity (red dashed). The smooth black curve is low-pass filtered curvature. The strongest refraction occurs when the ray crosses the ACC and equatorial current. Weaker refraction occurs as the ray transits the subtropical frontal zone.}
\label{validation}
\end{figure}

Fig. \ref{validation} compares the left and right hand sides of equation (\ref{Dysthe}) along a ray connecting the  southern source to the  receiver. The two curves coincide almost to within the line width. This validates the weak-current approximation used in (\ref{Dysthe}) and  shows that  rays are bent from great circles only where there is strong surface vorticity i.e., in localized current systems such as   the Antarctic Circumpolar Current (ACC), the subtropical frontal zone and the equatorial current system. The thick ray of Fig. \ref{fig5} undergoes strong refraction when crossing these three features, and then travels more-or-less on  great circles between these current systems. For example, after leaving the Southern Ocean the thick ray in Fig. \ref{fig5} is on a great circle headed away from the receiver. But refraction by a large equatorial eddy subsequently bends the ray onto a great circle passing through the  receiver.

We conclude that fluctuations in incident direction observed at the receiver are not the result of accumulation of many small random deflections. Instead, the model indicates that there are  two or three large deflections of a ray that are associated with major hydrographic features of the surface current system.  

\subsection{An extended source in the Southern Ocean}

\begin{figure}
\begin{center}
\includegraphics[width=100 mm]{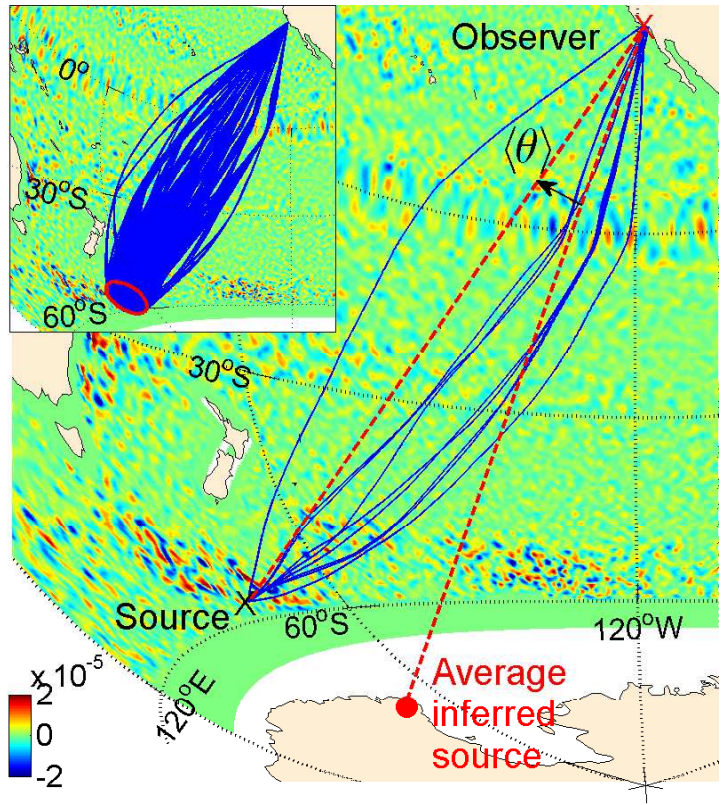}
\end{center}
\caption{
A bundle of rays connects the source to the receiver ($\lambda=500$m). Blue rays reach the receiver within 30 nautical miles. The red spot indicates the average inferred source. \textbf{Insert:} same analysis performed with an extended source of radius $4^\circ$, delimited by the red circle. The average inferred source coincides with the center of the circle within $2^\circ$ (not shown). This figure uses an azimuthal equidistant projection centered on the receiver and the color scale shows vertical  vorticity of OSCAR surface currents, $\xi$ in s$^{-1}$.}
\label{fig2} 
\end{figure}

For a given surface velocity field, several rays --- none of which are great circles ---  connect the source to the receiver.  An example is shown in Figure \ref{fig2}. We refer to the collection of rays that connect the source to the receiver as a ``multipath".

As we did with the directional data in Fig. \ref{2swellevents},  an  observer can construct the ``average inferred source" at the location of the mean position of the multiple inferred sources. This mean deflection is denoted as $\left< \theta \right>$, where the brackets denote averaging over all rays. For a point-source of swell, the average inferred source is usually significantly displaced from the actual source, with values of the order of $\pm 5^\circ$ for swell with initial wavelength $\lambda=500$m. An extreme example is presented in figure \ref{fig2}, where refraction by surface currents is so strong that the average inferred source is on land. However, real storms have a typical extension of several hundreds of kilometers and should not be considered to be point-sources. 

To determine the influence of this spatial extension, we model an extended source by a disk of radius $4^\circ$, or $440$ km, around a center located at latitude $57^\circ$S and longitude $172^\circ$ E. We assume a uniform density of incoherent point sources inside this disk. It is then easier to solve numerically the backward problem: we shoot rays ``backwards" from the receiver with a constant step of $10^{-4}$ radian in initial direction (approximately twenty seconds of arc). We keep only the rays that have a nonzero intersection with the swell-emitting disk in the Southern Ocean. To compute the direction of the average inferred source, we perform a weighted average of the angular directions  of the rays that intersect the extended source. The weight of each ray is equal to the length of its intersection with the swell-emitting disk: the contributions from incoherent sources inside the disk add up on each ray. The insert in figure \ref{fig2}  shows a typical ray pattern obtained with this extended source for swell with wavelength $\lambda=500$m. A wide bundle of rays connects the source to the receiver. We find that the average inferred source is close to the center of the actual source: contributions to the average deflection from the different points inside the emitting-disk average out to almost zero. Going from one 5-day frame of the OSCAR-data to the next one, the average deflection $\left< \theta \right>$ fluctuates weakly around zero, with a root-mean-square (rms) value of the order of $3^\circ$ for swell with $500$m initial wavelength: most often, the average inferred source is inside the swell-emitting disc. Assuming gaussian statistics for the average deflection $\left< \theta \right>$, such a low $3^{\circ}$ value of the rms fluctuations means that an average deflection of $10^\circ$ or higher has a probability lower than 0.04. This does not rule out the possibility of a large $\langle \theta \rangle$ occasionally occurring   because the  OSCAR data underestimate the actual vorticity of surface currents and because 0.04 is not 0:  if we analyze 25 events we might hope to see one example of a $10^{\circ}$ average deflection. This is consistent with the analysis from section \textbf{2}: mirages due solely to the effect of surface currents are  rare events. An additional ingredient seems therefore necessary to explain the more frequent occurrence of strong average deflections in the observations by Munk et al., together with the systematic displacement of the inferred sources towards the South of the actual storms. We return to this in section \textbf{6}.

\section{Fluctuations in the direction of swell}

We return now to the $\pm 10^{\circ}$ directional fluctuations in Fig. \ref{2swellevents}(c) and \ref{2swellevents}(d). Previously, to estimate the location of the source, we removed these fluctuations by averaging. But  in this section we investigate the hypothesis that the frequency dependence of directional fluctuations contains information about the strength of the surface vorticity field. 

Let us assume that, during a storm, surface wave packets are emitted somewhat randomly along the different rays that connect the storm to the receiver. These different wave packets will not reach the receiver at the same time, and thus the direction of the incident swell should fluctuate in time, with fluctuations of the order of typical values of $\theta$ visible in Fig. \ref{fig2}. Our hypothesis is that propagation of wave packets along the various component rays of  the multipath  in Fig. \ref{fig2}  results in the $\pm 10^{\circ}$ fluctuations in incident angle shown in  Fig. \ref{2swellevents}(c) and \ref{2swellevents}(d). To investigate this further we first characterize the fluctuations in direction induced by surface currents, using the twenty years of OSCAR data to perform a Monte Carlo computation. We then compare  this prediction to the pitch-and-roll buoy data.

\subsection{Directional fluctuations in model based on OSCAR currents}

\begin{figure}
\begin{center}
 \subfigure[]{
\includegraphics[width=72 mm]{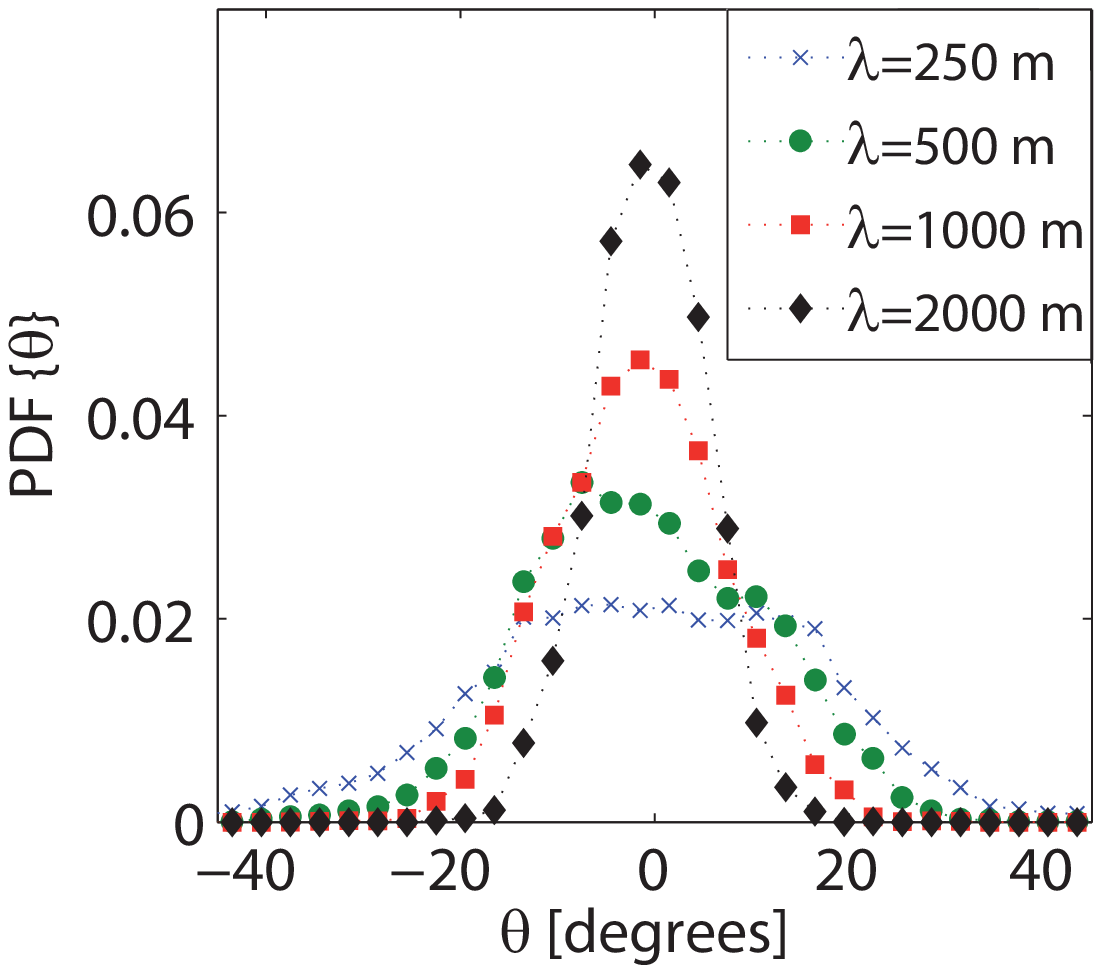}
}
 \subfigure[]{
\includegraphics[width=72 mm]{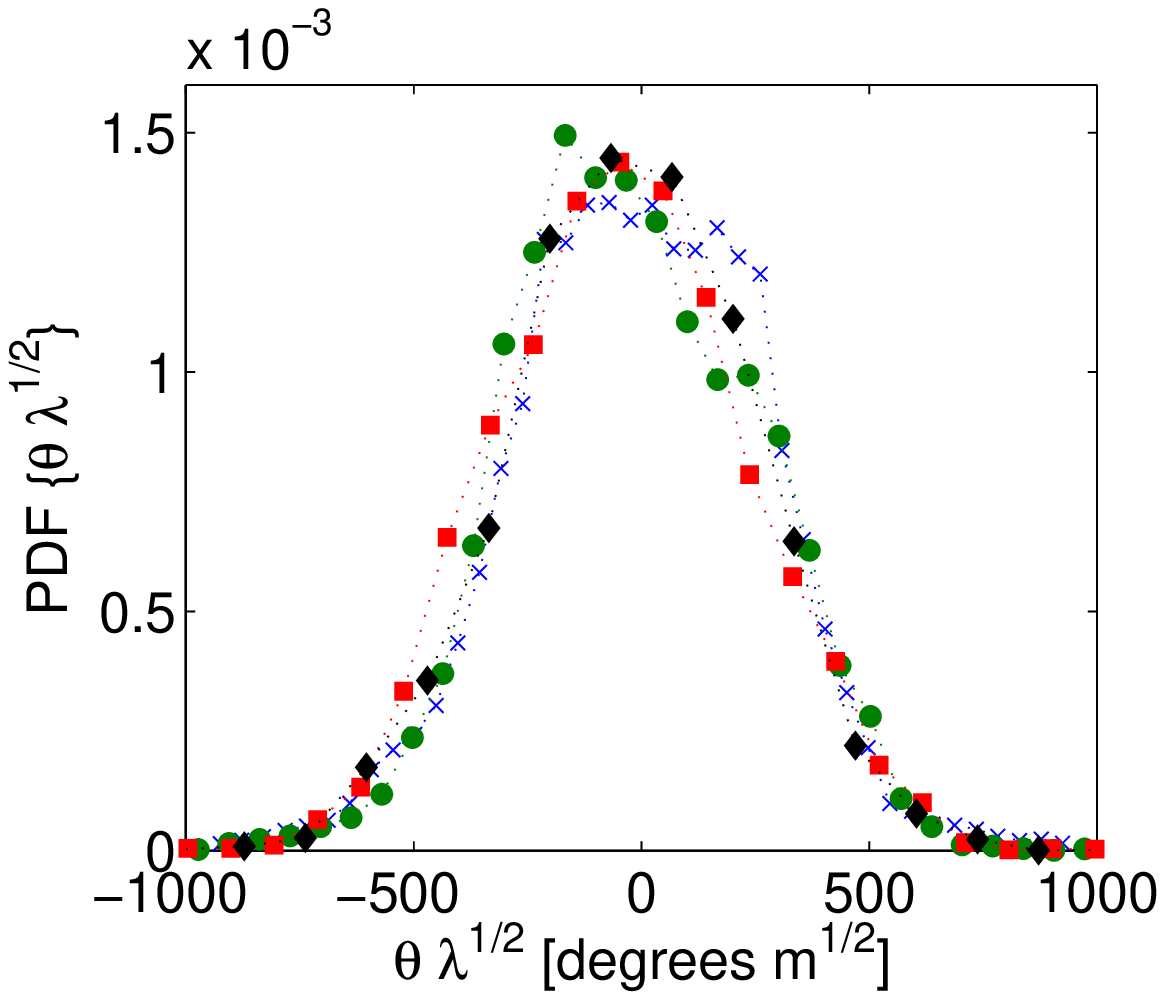}
}
\end{center}
\caption{(a) Probability density function of the deflection angle. The typical deflection is around $10^{\circ}$, which is consistent with the measurements of Munk \textit{et al.} and the observation of section \textbf{2}. Waves with larger wavelength are faster and less refracted. (b) The four PDFs collapse onto a  master PDF when considered a function of $\theta \sqrt{\lambda}$.}
\label{fig4} 
\end{figure}

To gather statistics on the deflection of the rays, we repeat the point-source analysis of section \textbf{4} for each 5-day OSCAR surface velocity field recorded between October 1992 and October 2011. From this extensive simulation, we compute the probability density function (PDF) of the deflection angle,  $\theta$, which  is  defined as the angle between the ray at arrival and the great-circle route. This PDF of $\theta$ is shown in figure \ref{fig4}(a)  using four different  values of initial swell wavelength $\lambda$. The root-mean-square (rms) deflection is around $17^\circ$ for wavelength $\lambda=250$m. The rms deflection decreases with increasing wavelength. Indeed, longer waves are faster and thus less refracted according to the curvature equation (\ref{Dysthe}): when  considered as a function of $\theta \sqrt{\lambda}$, the four PDFs collapse onto a single master curve: see Fig. \ref{fig4}(b). This collapse indicates an rms deflection angle proportional to the frequency of the swell,
\begin{equation}
\theta_{\mbox{rms}} = 218 f \, , \label{eqthetarms}
\end{equation}
with $\theta_{\mbox{rms}}$ in degrees and $f$ in Hertz.


\begin{figure}
\begin{center}
\includegraphics[width=100 mm]{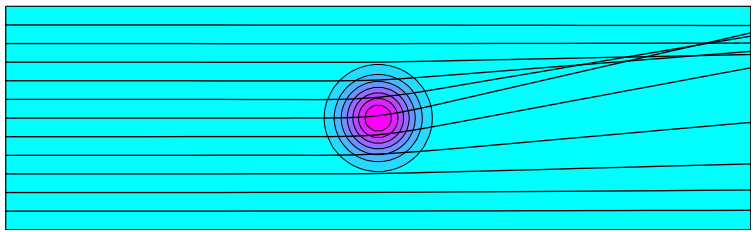}\\
\end{center}
\caption{Bending of a beam of parallel rays by a strong Gaussian vortex $\xi = \xi_0 \exp(- (r/\ell)^2)$, with $\xi_0/\sqrt{gk} = 5/\pi$ and $\ell k  = 40$. In this illustration the vortex is unrealistically strong in order to show the ray pattern; realistic vortices produce much smaller deflections. }
\label{vorty} 
\end{figure}

The order of magnitude of typical  deflection $\theta$ can be understood as follows: according to the curvature equation (\ref{Dysthe}), when a ray crosses an eddy with size $\ell$ and typical vorticity $\xi_0$ (see Fig. \ref{vorty}), the direction of propagation is deflected by an angle
\begin{align}
\theta &\sim \frac{\xi_0 \ell}{v_g} \, \\
&= 4 \pi \frac{\xi_0 \ell}{g} \, f \, , \label{ODG}
\end{align}
where  $v_g= g/(4 \pi f)$ has been used.


Out of the three current systems that refract the swell, the ACC has the most intense vorticity (large $\xi_0$) whereas the equatorial current has the largest eddies (large $\ell$). The subtropical frontal zone has weaker eddies with small sizes and can be neglected for this rough estimate. Because the ACC is close to the source and far from the receiver, the ACC has a smaller impact on the deflection $\theta$ than the equatorial current system: even after strong refraction in the ACC, rays escape this first current system rather close to the source point. If the ACC  were the only vortical refractor between source and receiver, then the observer would make only a small error in inferring the position of the source. But the equatorial current system is close to the receiver. Equatorial eddies bend some rays so that they hit the receiver, which results in large values of $\theta$. Hence an order of magnitude of $\theta$ is given by the angular shift due to a few, or even one,  eddy in the equatorial current system. With $\lambda=500$m, and using $\ell=400$km and $\xi_0 = 10^{-5} \mbox{s}^{-1}$ for a typical eddy in the equatorial current system, one obtains from \eqref{ODG} $\theta \simeq 9^\circ$. 

Although Kenyon's intuition that surface currents refract the swell was correct, he focused on the time-averaged ACC and neglected both the equatorial current and the role of the eddies. The latter are crucial to explain the directional fluctuations due to OSCAR currents. Indeed, to determine the role of the average currents, we time-averaged the twenty years of OSCAR data before simulating the same point-source as in section \textbf{4}a: the time-averaged surface currents produce a negligible deflection $|\theta|$ of the rays (typically $1.5^\circ$ for $\lambda=500$ m), which shows that the strong fluctuations visible in figure \ref{fig4} are due to the small-scale eddies, particularly their intense vorticity,  and not to the time-averaged currents.

\subsection{Directional fluctuations in the pitch-and-roll buoy data}

\begin{figure}
\begin{center}
\includegraphics[width=80 mm]{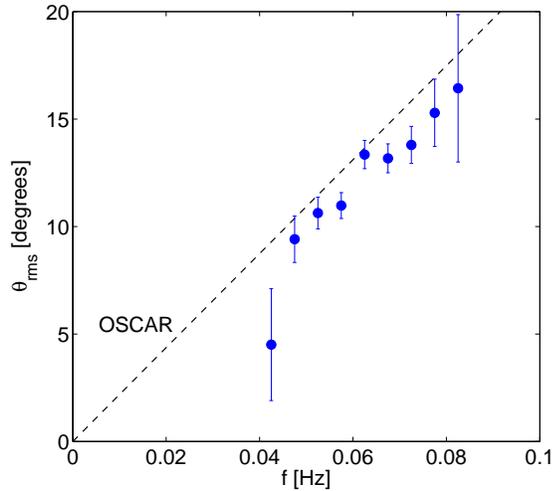}\\
\end{center}
\caption{Root-mean-square fluctuations in incoming swell direction measured by a pitch-and-roll buoy, averaged over 18 storms, as a function of swell frequency $f$. Error bars are evaluated using the square root of the number of values in each frequency bin. The dashed-line is the result (\ref{eqthetarms}) from the analysis of OSCAR data.}
\label{thetarmsvsf_allrays} 
\end{figure}

To gather  statistics on the fluctuations of $\theta$ in the buoy data shown in Fig. \ref{2swellevents}, we analyzed 18 swell events recorded between 2004 and 2007. These 18 events were selected  because the signal was particularly clean i.e., there was no evidence of multiple sources in any of these 18 events. For each swell event, we removed the mean value and the linear trend in the incident angle signal. We consider the remaining fluctuating $\theta(t)$ as a function of the peak frequency $f(t)$ instead of time. We divide the signal into small bins in $f$, and we compute the root-mean-square $\theta$ for each bin. The result is displayed in Fig. \ref{thetarmsvsf_allrays}, together with the predictions from the analysis of the OSCAR data. The rms value of $\theta$ indeed shows a linear dependence with frequency, as predicted for refraction by surface currents. More surprisingly, the prefactor of the linear law is very well captured by the OSCAR data analysis.  We were expecting to underestimate the effect of surface currents using the coarse OSCAR data so the agreement in Fig. \ref{thetarmsvsf_allrays} might be fortuitous. But the orders of magnitude are definitely compatible.


The good comparison in Fig.  \ref{thetarmsvsf_allrays} is also  hostage to errors in  directional measurements at NOAA station 46086. A significant uncertainty is that the accelerometers on NDBC 3m discus buoys (such as buoy 46086)  are known to be a little noisy, or at least they are noisier than accelerometers on Datawell Directional Waveriders \citep{BOR}. The discus-buoy noise is known to bias estimates of spread, but  does not  influence estimates of mean direction \citep{BOR}. To further allay these concerns, the November 2007 case study has been repeated by Sean Crosby (personal communication) using measurements made by a Datawell buoy  deployed at NOAA NDBC station 46232 (32.530$^{\circ}$N and 117.431$^{\circ}$W). Datawell buoys have the advantage of measuring accurately both the mean direction and the spread of the swell, using 26-minute-long samples. The Datawell directional measurements are consistent with the  NDBC results:  directional spread increases in time as the peak frequency increases. The fluctuations in incident angle (or ``mean direction" signal) are greater than the Datawell noise level, and their magnitude increases between the beginning and the end of the swell event.  These  results are consistent with our  hypothesis that some part of the observed  directional  fluctuations is due to  ray bending by surface currents.

\section{Discussion and conclusion}

\begin{figure}
\begin{center}
 \subfigure[]{
\includegraphics[width=100 mm]{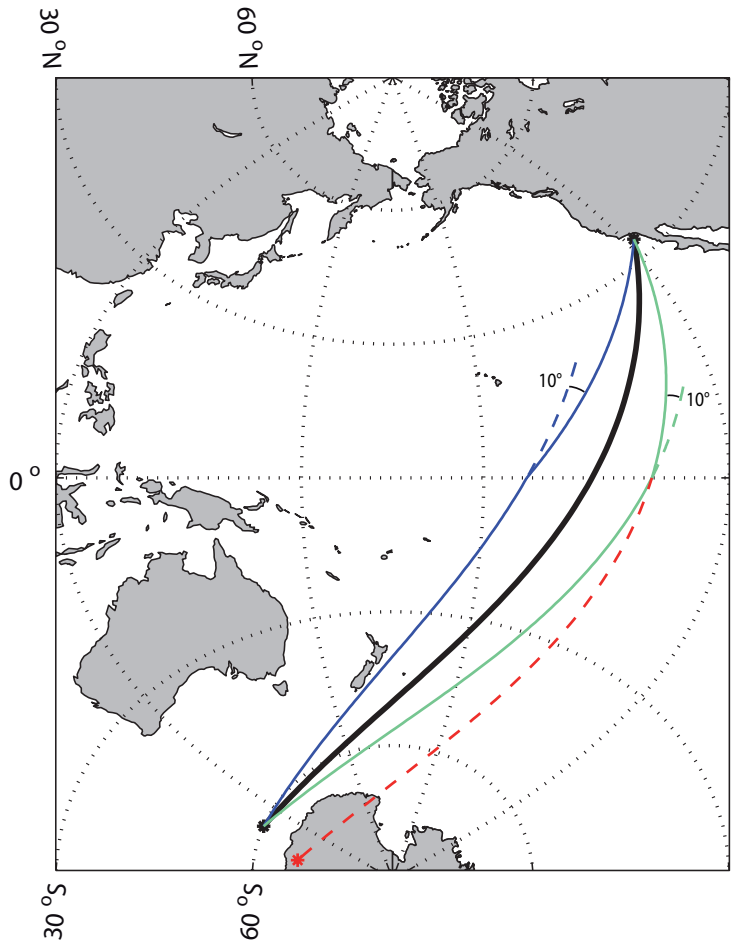}
}
 \subfigure[]{
\includegraphics[width=100 mm]{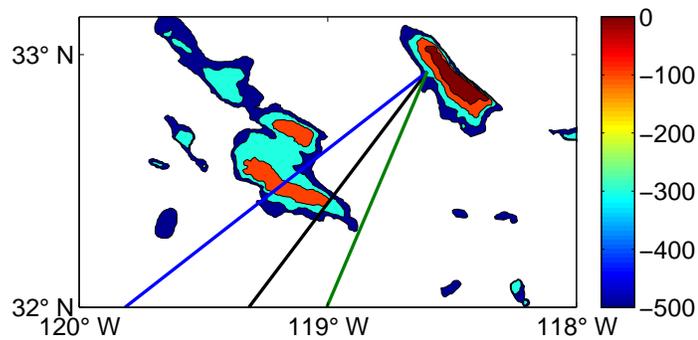}
}
\end{center}
\caption{Interplay between surface currents and shallow bathymetry. (a) The black line is the great-circle route. The blue and green rays connect the source to the receiving station from Munk et al. (1963). They are deflected by $\pm 10^\circ$ by mesoscale vorticity at the equator. The Northern ray hits Cortez bank, and does not reach the receiver. An observer sees only the Southern ray and great-circle backtracking puts the source on Antarctica (dashed red line). (b) Local bathymetry in meters near San Clemente island. Cortez bank blocks the Northern ray.}
\label{bent_rayz} 
\end{figure}

Because of refraction by surface currents, a storm and a receiving station are  connected not only by the great-circle route, but by a multipath: a bundle of rays with an angular width which is much larger than the angular width of the storm. Wave packets travel on these many rays before reaching the receiver, which leads to strong temporal fluctuations in the incoming direction measured at the receiver. The root mean square fluctuations in the directional signal at pitch-and-roll buoy 46086 are consistent with predictions using ray-tracing through the mesoscale vorticity of the OSCAR dataset.

Most often, the fluctuations average out in time and the mean direction measured by the buoy over several days coincides with the actual direction of the storm. However, mirages do occur. These  rare events are observed when most of the swell from a single storm is deflected in the same direction. The mirage effect can be greatly enhanced by topography: if  topographic features close  to the great-circle route obstruct part of the ray bundle, then the average direction of the incoming swell can be very different from the direction of the storm.

The historical case of \citet{Munk63} probably results from such an interplay between surface currents and topography, summarized following Munk (2013) in the simplified schematic Fig. \ref{bent_rayz}. We represent the great-circle route between a storm and the receiver off San Clemente island, together with the extreme rays experiencing a refraction of respectively $+10^\circ$ and $-10^\circ$ by mesoscale vorticity in the equatorial current. Now close to the receiver is Cortez bank, a shallow bathymetric structure that selectively blocks rays that travel North of the great-circle route without affecting rays that approach from the  South. An observer therefore measures mostly wave packets coming from South of the great-circle route, and infers a source on Antarctica.



Satellite observations provide modern confirmation of long-range propagation of ocean swell \citep{Heimbach,Collard}. But the source-location problem remains an issue: recent inferences of storm sources  follow \cite{Munk63} and backtrack along great circles. This  ignores refraction by currents   and the resulting increase in the width of the swell beam. It is interesting that through the curvature formula \eqref{Dysthe}, vorticity $\xi$ emerges as a key environmental variable that controls refraction of surface gravity waves.  NASA's Surface Water and Ocean Topography satellite promises to greatly improve resolution of mesoscale vorticity \citep{Fu,FuFerrari}: better wave forecasts might be an unexpected outcome of this mission.

\bigskip
\emph{Acknowledgments.}
This research was supported by the  National Science Foundation under OCE10-57838; BG was partially supported by a Scripps Postdoctoral Fellowship. We thank  Ryan Abernathey, Fabrice Ardhuin, Oliver B\"uhler, Fabrice Collard, Sean Crosby, Falk Feddersen and particularly Walter Munk,  for many conversations and help with this problem.

The ocean surface current data used in this study was provided by the  OSCAR Project Office. The significant wave height  data were obtained from the ECMWF data server. The pitch-and-roll buoy data is from  the National Data Buoy Center, which is operated by the National  Oceanic and Atmospheric Administration.   

\clearpage
\bibliographystyle{JMR}
\bibliography{billzBib.bib}

\listoffigures

\end{document}